\documentclass[english,prl,twocolumn,reprint]{revtex4-1}
\usepackage[T1]{fontenc}
\usepackage[latin9]{inputenc}
\usepackage{amsmath}
\usepackage{graphicx}
\usepackage{esint}

\makeatletter

\providecommand{\tabularnewline}{\\}

\usepackage{graphicx}
\DeclareRobustCommand{\qp}{\text{\reflectbox{p}}\mkern-3mu\text{p}}
\DeclareRobustCommand{\rhoqp}{\text{\reflectbox{P}}\mkern-4mu\text{P}}

\makeatother

\usepackage{babel}
\begin{document}

\title{Integral formulation of the quantum mechanics in the phase space}

\author{Tom\'{a}\v{s} Zimmermann }
\email{tomas.zimmermann@uzh.ch}

\affiliation{Department of Chemistry, University of Zurich, Switzerland}
\begin{abstract}
A formulation of quantum mechanics is introduced based on a $2D$-dimensional
phase-space wave function $\text{\reflectbox{\text{p}}}\mkern-3mu\text{p}\left(q,p\right)$
which might be computed from the position-space wave function $\psi\left(q\right)$
with a transformation related to the Gabor transformation. The equation
of motion for conservative systems can be written in the form of the
Schr\"{o}dinger equation with a $4D$-dimensional Hamiltonian with
classical terms on the diagonal and complex off-diagonal couplings.
The Hamiltonian does not contain any differential operators and the
quantization is achieved by replacing $q$ and $p$ with $2D$-dimensional
counterparts $\left(q+q'\right)/2$ and $\left(p+p'\right)/2$ and
by using a complex-valued factor $e^{i\left(q\cdot p'-q'\cdot p\right)/2}$
in phase-space integrals. Despite the fact that the formulation increases
the dimensionality, it might provide a way towards exact multi-dimensional
computations as it may be evaluated directly with Monte-Carlo algorithms.
\end{abstract}
\maketitle
Phase-space representations have been sought after since the early
days of the quantum mechanics. Initially, several phase-space distributions
based on the density matrix have been found. The most prominent among
them is undoubtedly the Wigner function \cite{Wigner1932,Moyal1949,Groenewold1946},
but other functions such as Kirkwood-Rihaczek distribution \cite{Kirkwood1933},
Husimi function \cite{Husimi1940} or Glauber-Sudarshan $P$ representation
\cite{Glauber1963a,Sudarshan1963} are in use. The state-functions
underlying these formulations have properties of the density matrix,
most notably they are not additive and have coherences between different
parts of the corresponding wave packet. Since the equations of motion
of these distributions are typically more complicated than the Schr\"{o}dinger
equation, they are rarely used in (fully quantum-mechanical) numerical
simulations. On the other hand, the Wigner function (together with
the Husimi distribution) often serves as the base for semiclassical
approximations due to its straightforward connection with the classical
mechanics \cite{Heller1976,Miller2001,Zimmermann2014,Zimmermann2018}.

Another attempt to represent the quantum mechanics in the phase space,
which was also found in the early days, is based on the expansion
of the position-space wave function into the basis of Gaussian wave
packets \cite{Neumann1931,Perelomov1971}. Even though the Gaussian
wave packet is also defined in the position space, it has a well-defined
mean momentum and the approach might be though of as an indirect way
of introducing the wave-function into the phase space. Due to its
favorable scaling with the system size, the approach is often used
in numerical simulations \cite{Shalashilin2001,Shimshovitz2012}. 

More recently, direct phase-space formulations with functions based
on the wave function have been found \cite{Torres-Vega1993,Moller1997,Watson2011,Campos2016,Zuniga-Segundo2016},
mostly based on the Gabor transform of the position-space wave function.
Similarly to the wave function, these functions are additive and their
square, which is equal to the Husimi function \cite{Torres-Vega1993,Moller1997,Zuniga-Segundo2016},
might be interpreted as the phase-space probability. Nonetheless,
the equations of motion of the functions directly based on the Gabor
transformation contain differential operators which complicate their
application. The goal of the present work is to introduce the phase-space
formulation of the quantum mechanics which shares many properties
with the aforementioned wave function\textendash based formulations
and which is governed by the equation of motion which does not contain
any differential operators.

\begin{figure*}
\begin{tabular}{ccc}
\includegraphics[width=0.33\textwidth]{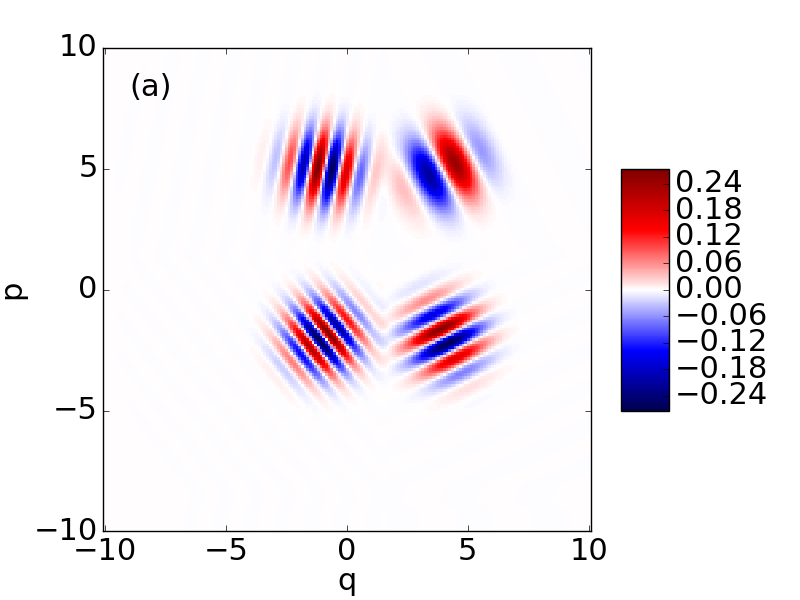} & \includegraphics[width=0.33\textwidth]{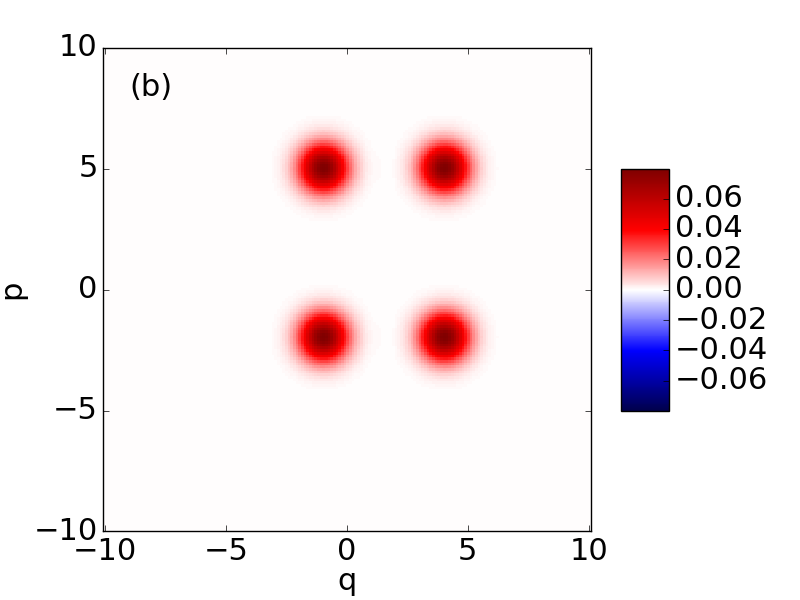} & \includegraphics[width=0.33\textwidth]{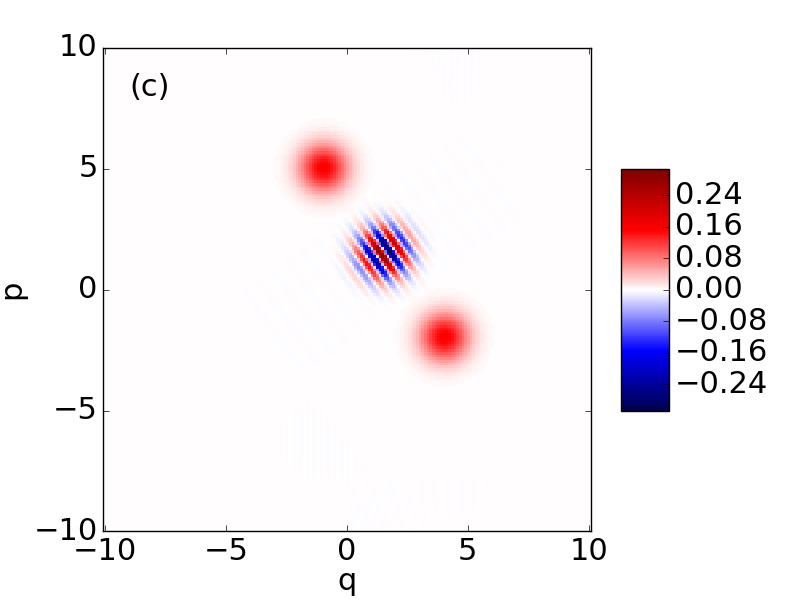}\tabularnewline
\end{tabular}

\caption{Density-matrix based phase-space representations of the quantum state
composed of two Gaussian wave functions $z_{1}$ and $z_{2}$ centered
at $\left(q,p\right)=\left(4,-2\right)$ and $\left(-1,5\right)$:
$\psi\left(q\right)=N\left[z_{1}\left(q;4,-2\right)+z_{2}\left(q;-1,5\right)\right],$
$N$ is the normalization constant. (a) $\text{Re}\left[P\left(q,p\right)\right]$,
(b) $\left|P\left(q,p\right)\right|^{2}$ , and (c) $W\left(q,p\right)$.
Note that the coherences between $z_{1}$ and $z_{2}$ in $P\left(q,p\right)$
are located at the corners of a phase-space rectangle whereas in $W\left(q,p\right)$
the coherences are in the middle of the phase-space line connecting
$z_{1}$ and $z_{2}$. \label{fig:two_gauss_wigner}}
 
\end{figure*}

Following the historical development, probably the most straightforward
way to obtain a phase-space description of a quantum state starts
with the density matrix $\rho$ which may be expressed in the position-space
as $\rho\left(q',q\right)=\psi\left(q'\right)\psi^{*}\left(q\right)$.
A phase-space density matrix $P\left(q,p\right)=\psi\left(p\right)\psi^{*}\left(q\right)$
might be computed by simply transforming $\psi\left(q'\right)$ into
the momentum-space representation $\psi\left(p\right)$
\begin{equation}
P(q,p)=\left(2\pi\right)^{-\frac{D}{2}}\int d^{D}q'\,\psi\left(q'\right)\psi^{*}\left(q\right)\,e^{-iq'\cdot p},\label{eq:P_transform}
\end{equation}
where $D$ is the number of dimensions. (All formulas are in atomic
units.) Note that $P$ is closely related to the Kirkwood-Rihacek
distribution \cite{Kirkwood1933}, which differs only by a complex
factor $e^{-ipq}$. In a more elaborate approach, $q$ and $q'$ are
first transformed to the average and difference variables and the
difference variable is transformed into the momentum-space resulting
in the Wigner function $W(q,p)$ 
\begin{equation}
W(q,p)=\left(2\pi\right)^{-D}\int d^{D}\xi\,\psi\left(q+\xi/2\right)\psi^{*}\left(q-\xi/2\right)\,e^{-i\xi\cdot p}.\label{eq:Wigner_transform}
\end{equation}
Both $P$ and $W$ are based on the density matrix and are quadratic
with respect to $\psi$. Therefore they contain coherences between
components of $\psi$. As can be seen in Fig.~\ref{fig:two_gauss_wigner}
which plots $P$ and $W$ for two coherent states 
\begin{equation}
z\left(q;q_{0},p_{0}\right)=\pi^{-\frac{D}{4}}\exp\left[ip_{0}\cdot\left(q-q_{0}\right)-\frac{\left(q-q_{0}\right)^{2}}{2}\right],\label{eq:coherent_state}
\end{equation}
in $W(q,p)$ the coherences are located directly in between the Wigner
functions of the components of the underlying $\psi$, whereas in
$P\left(p,q\right)$ they are in the opposite corners of the rectangle
which has $P$s corresponding to components of $\psi$ located on
the other diagonal. This pattern suggests that the coherences are
in a certain sense trivial and can be computed from simpler underlying
phase-space function $\qp\left(q,p\right)$ which would be linear
with respect to $\psi$. 

Analysis of $\psi$ and $P$ corresponding to Gaussian wave packets
and simple functions such as $\psi\left(q\right)=sin(p_{0}q)=i/2\left(e^{-ip_{0}q}-e^{ip_{0}q}\right)$
suggests that the most natural transformation to compute $P$ from
smoother ``coherence-free'' phase-space wave function $\qp\left(q,p\right)$
should have the form
\begin{equation}
P(q,p)=\left(4\pi\right)^{-\frac{D}{2}}\iint d^{D}q'd^{D}p'\rhoqp\left(q',p,q,p'\right)e^{-i\frac{q'\cdot p+q\cdot p'}{2}},\label{eq:z2p}
\end{equation}
where the 4$D$ dimensional density matrix $\rhoqp\left(q',p,q,p'\right)=\qp\left(q',p\right)\qp^{*}\left(q,p'\right)$.
The momentum- and position-space $\psi$ can then be computed as
\begin{align}
\psi(q) & =\left(2\sqrt{\pi}\right)^{-\frac{D}{2}}\int d^{D}p'\,\qp\left(q,p'\right)\,e^{i\frac{q\cdot p'}{2}}\label{eq:z2psi_q}\\
\psi(p) & =\left(2\sqrt{\pi}\right)^{-\frac{D}{2}}\int d^{D}q'\,\qp\left(q',p\right)\,e^{-i\frac{q'\cdot p}{2}}.\label{eq:z2psi_p}
\end{align}

\begin{figure*}
\begin{tabular}{ccc}
\includegraphics[width=0.33\textwidth]{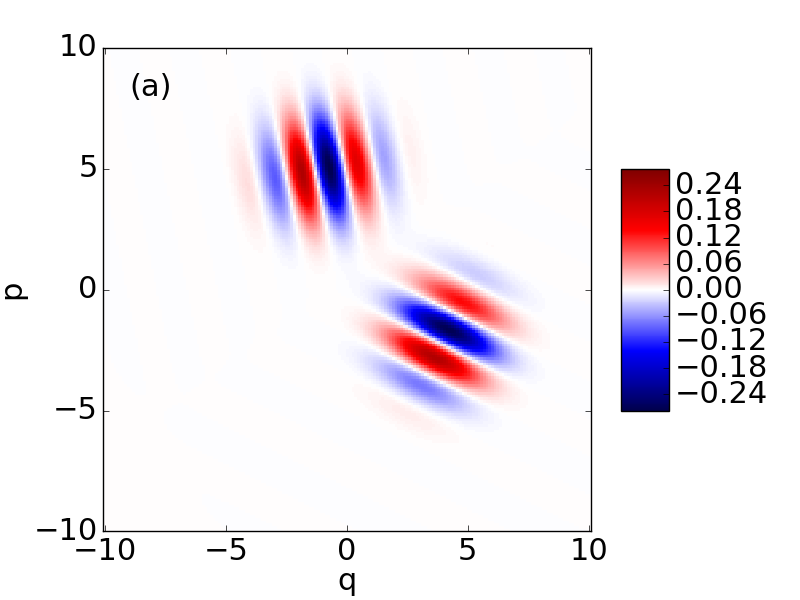} & \includegraphics[width=0.33\textwidth]{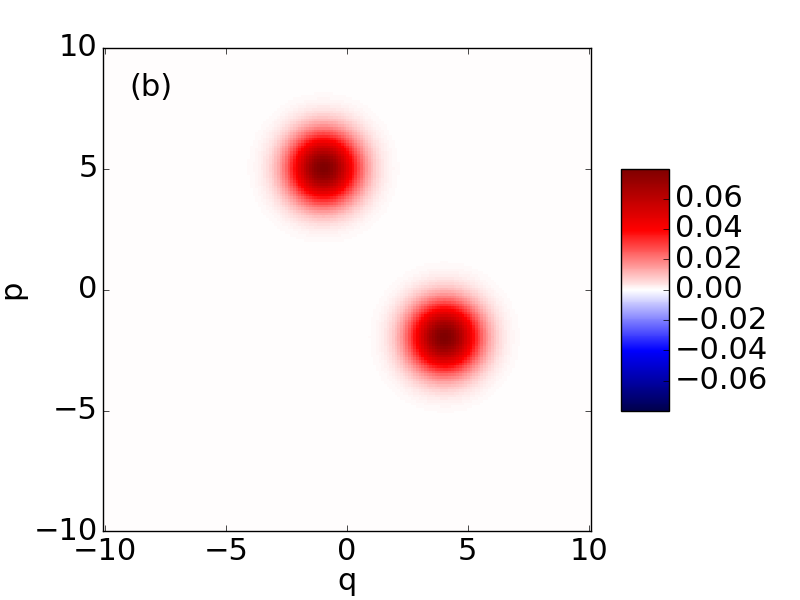} & \includegraphics[width=0.33\textwidth]{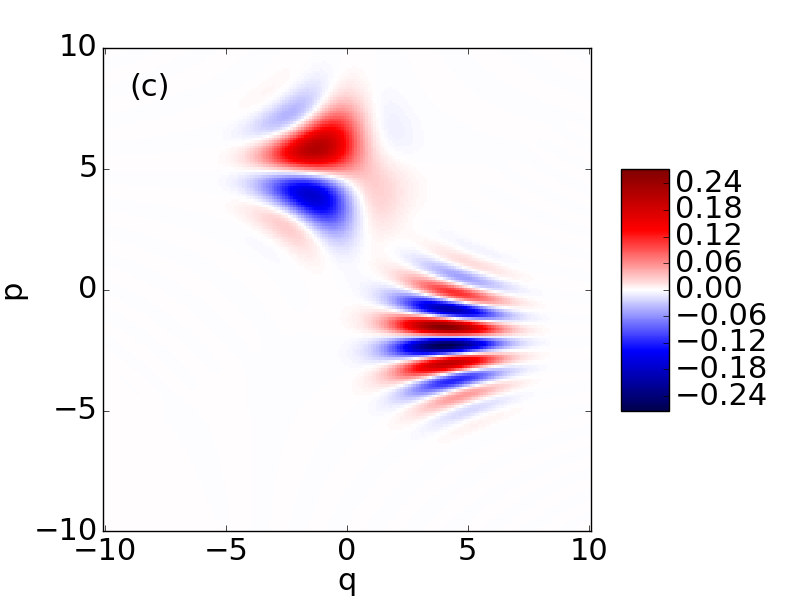}\tabularnewline
\end{tabular}

\caption{$\qp\left(q,p\right)$ of the quantum state composed of two Gaussian
wave functions $z_{1}$ and $z_{2}$ centered at $\left(q,p\right)=\left(4,-2\right)$
and $\left(-1,5\right)$. (a) $\text{Re}\left[\qp\left(q,p\right)\right]$,
(b) $\left|\qp\left(q,p\right)\right|^{2}$, (c) $\text{Re}\left[G\left(q,p\right)\right]$,
where $G\left(q,p\right)=\left(2\pi^{\frac{3}{2}}\right)^{-\frac{D}{2}}\int d^{D}q'\,e^{\frac{-\left(q-q'\right)^{2}}{2}}\psi\left(q'\right)e^{-iq'\cdot p}$
is computed with the Gabor transform without the complex prefactor
$e^{i\frac{q\cdot p}{2}}$. Note that $\qp\left(q,p\right)$ is substantially
more diffuse than density-matrix based functions, that there is no
coherence between $z_{1}$ and $z_{2}$, and that the phase of $\qp\left(q,p\right)$
varies slowly (due to a factor of $1/2$). Functions $\qp\left(q,p\right)$
and $G\left(q,p\right)$ differ only by a phase. \label{fig:two_gauss_qp}}
\end{figure*}

It may be shown (see Appendix) that Eqs.~(\ref{eq:z2p})-(\ref{eq:z2psi_p})
are satisfied by $\qp\left(q,p\right)$ computed from $\psi(q)$ with
a Gabor-like transform 
\begin{equation}
\qp\left(q,p\right)=\left(2\pi^{\frac{3}{2}}\right)^{-\frac{D}{2}}e^{i\frac{q\cdot p}{2}}\int d^{D}q'\,e^{\frac{-\left(q-q'\right)^{2}}{2}}\psi\left(q'\right)e^{-iq'\cdot p}.\label{eq:psi_q2z}
\end{equation}
An example of $\qp$ corresponding to the sum of two Gaussian wave
packets is shown in Fig.~\ref{fig:two_gauss_qp}. Note that, in contrast
to phase-space functions based on the density operator such as Wigner
function or Kirkwood-Rihaczek function, the coherence term is missing.

To complement the set of transformations between $\qp\left(q,p\right)$
and density-matrix based phase-space functions, it was found and numerically
verified that for 1-$D$ systems the Wigner transformation might be
computed from $\qp\left(q,p\right)$ with the formula\begin{widetext}

\begin{equation}
W_{1D}(q,p)=\frac{1}{\pi}\iint d\xi d\upsilon\,\qp\left(q-\xi,p-\upsilon\right)\qp^{*}\left(q+\xi,p+\upsilon\right)\,e^{-i\left|\eta\times\zeta\right|\cdot n},\label{eq:z2w}
\end{equation}
\end{widetext}where $\eta=\left(q,p\right)$, $\zeta=\left(\xi,\upsilon\right)$,
$n$ is the unit vector orthogonal to the $\left(q,p\right)$ plane,
and $\left|\eta\times\zeta\right|\cdot n=\left|\eta\right|\left|\zeta\right|\sin\left(\beta\right)$,
where $\beta$ is the angle between vectors $\left(q,p\right)$ and
$\left(\xi,\upsilon\right)$ measured from $\left(q,p\right)$ counterclockwise.

Due to the linearity of the transformation~(\ref{eq:psi_q2z}), the
function $\qp\left(q,p\right)$ shares many properties with $\psi\left(q\right)$
such as the additivity. The diagonal of the density matrix $\rhoqp\left(q,p,q,p\right)$
is equal to the Husimi function \cite{Torres-Vega1993,Moller1997}
and its trace is for pure states equal to unity
\begin{equation}
I_{\qp,\qp}=\iint dq^{D}dp^{D}\,\rhoqp\left(q,p,q,p\right)=1.\label{eq:probability_integral}
\end{equation}
This suggests that $\rhoqp\left(q,p,q,p\right)$ might be interpreted
as the phase-space probability. In addition, $\qp\left(q,p\right)$
satisfies the phase-space identity transformation (for the proof see
the Appendix)
\begin{equation}
\qp\left(q,p\right)=\left(4\pi\right)^{-D}\iint d^{D}q'd^{D}p'\,\qp\left(q',p'\right)e^{i\frac{q\cdot p'-q'\cdot p}{2}}.\label{eq:qp2qp}
\end{equation}

Having established the connection between $\psi$ and $\qp$, the
next step is to determine the equation of motion for conservative
systems with the Hamiltonian in the form $\mathbf{\hat{H}=\hat{\mathbf{T}}}+\hat{\mathbf{V}}$.
The non-relativistic conservative evolution of $\psi\left(q\right)$
is governed by the Schr\"{o}dinger equation which might be written
as 

\begin{equation}
\partial_{t}\psi\left(q\right)=-i\int d^{D}q'\,H(q,q')\psi\left(q'\right)\label{eq:tdse}
\end{equation}
with the diagonal Hamiltonian
\begin{equation}
H(q,q')=\delta_{qq'}\frac{-\partial_{q}^{2}}{2m}+\delta_{qq'}V\left(q\right).\label{eq:position_space_Hamiltonian}
\end{equation}
 Analogously, the equation of motion for $\qp$ might be written as
(for the proof see the Appendix)\begin{widetext}
\begin{equation}
\partial_{t}\qp\left(q,p\right)=-i\left(4\pi\right)^{-D}\iint d^{D}q'd^{D}p'\,e^{i\frac{q\cdot p'-q'\cdot p}{2}}H(q,p,q',p')\qp\left(q',p'\right),\label{eq:prop}
\end{equation}
 \end{widetext}where 
\begin{equation}
H\left(q,p,q',p'\right)=\frac{1}{2m}\left(\frac{p+p'}{2}\right)^{2}+V\left(\frac{q+q'}{2}\right).\label{eq:4D_Hamiltonian}
\end{equation}
Note that the Hamiltonian multiplied with the phase-space complex
prefactor $e^{i\left(q\cdot p'-q'\cdot p\right)/2}$ is Hermitian
with respect to the interchange $\left(q,p\right)\longleftrightarrow\left(q',p'\right)$.
For a diagonal element $\left(q,p\right)=\left(q',p'\right)$, the
complex factors in Eq.~(\ref{eq:prop}) cancel out and $\qp\left(q,p\right)$
is multiplied directly with the classical Hamiltonian
\begin{equation}
H(q,p)=\frac{p^{2}}{2m}+V\left(q\right).\label{eq:classical_Hamiltonian}
\end{equation}
For $\left(q,p\right)\neq\left(q',p'\right)$, complex off-diagonal
terms couple two distant phase-space points. 

The mean value of an operator $\hat{\mathbf{O}}$ which is the function
of only $q$, or $p$, or their sum can be computed as (see the Appendix for the proof)
\begin{eqnarray}
\left\langle \mathbf{\hat{O}}\right\rangle  & = & \left(4\pi\right)^{-D}\iiiint d^{D}q'd^{D}p'd^{D}q''d^{D}p''\,e^{i\frac{q''\cdot p'-q'\cdot p''}{2}}\nonumber \\
 &  & \qp^{*}\left(q'',p''\right)O\left(\frac{q'+q''}{2},\frac{p'+p''}{2}\right)\qp\left(q',p'\right).\label{eq:means}
\end{eqnarray}
(Operators which contain products of $q$ and $p$ are not considered
here.) Similarly to the equation of motion, Eq.~(\ref{eq:means})
is the phase-space analog of the expression for the mean values of
operators in terms of the wave function $\psi$.

The equation of motion $\eqref{eq:prop}$ together with the formula
for mean values (\ref{eq:means}) suggest that, at least for conservative
systems, the quantization of the classical phase space may be achieved
by increasing the dimensionality of the Hamiltonian to $4D$, substituting
$q\rightarrow\left(q+q'\right)/2$ and $p\rightarrow\left(p+p'\right)/2$,
and using the phase-space integration factor $e^{i\left(q\cdot p'-q'\cdot p\right)/2}$. 

Although, the $\qp$ formulation actually increases the computational
effort by increasing the dimensionality introducing an explicitly
non-local Hamiltonian (in the sense that it couples every phase-space
point with every other phase-space point) it also offers several advantages
in comparison with the traditional formulation of the quantum mechanics
based on differential operators. Most notably, Eqs.~(\ref{eq:prop})
and (\ref{eq:means}) allow for direct Monte-Carlo evaluation sampling
the absolute value of $\left|\qp\left(q,p\right)\right|$ or $\left|\rhoqp\left(q,p,q',p'\right)\right|$,
so that a Monte Carlo based approach may be used for the evaluation
of properties as well as for propagation. Even though the complex
form of the equations will undoubtedly complicate the convergence,
the approach may provide a path towards calculations for highly-dimensional
systems. The phase-space identity transform (\ref{eq:qp2qp}) is especially
appealing in the context of Monte-Carlo methods as it provides a way
to directly compute $\qp\left(q,p\right)$ from the set of known values
$\qp\left(q',p'\right)$ via Monte-Carlo integration without interpolation.
The Monte-Carlo approach will be a topic of our subsequent publication.

In addition, the $\qp$ formulation might be useful in evaluating
complex functional forms of operators of $q$ and $p$. For example,
the relativistic expression for the energy
\begin{equation}
E(q,p)=\sqrt{c^{2}p^{2}+\left(m_{0}c^{2}\right)^{2}}+V\left(q\right)\label{eq:dirac}
\end{equation}
might be evaluated directly and it will be interesting to see whether
it might be used in Eq.~(\ref{eq:prop}) instead of the non-relativistic
Hamiltonian. The relativistic quantum mechanics might therefore represent
an interesting area of application of the $\qp$ formulation.

\begin{figure}
\begin{tabular}{c}
\includegraphics[width=1\columnwidth]{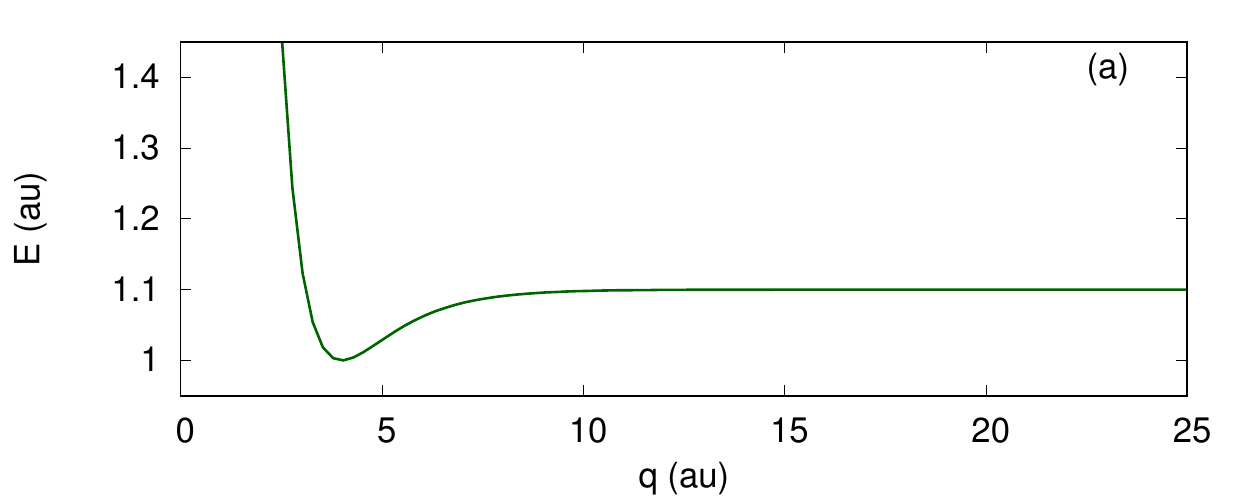}\tabularnewline
\includegraphics[width=1\columnwidth]{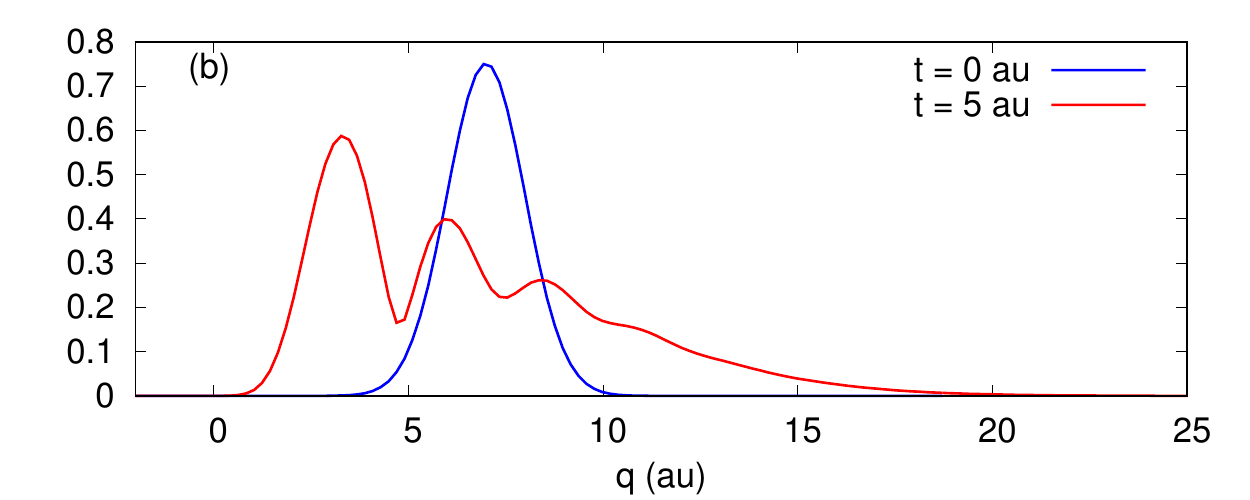}\tabularnewline
\includegraphics[width=1\columnwidth]{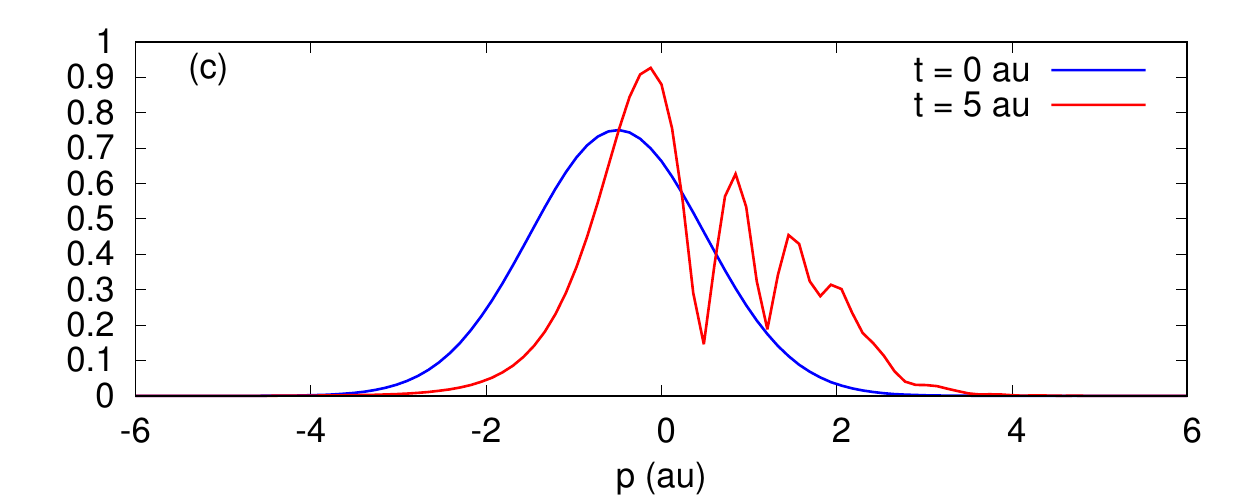}\tabularnewline
\end{tabular}

we arrive\caption{(a) Morse potential used in the numerical example. (b) initial ($t=0$
au) and final ($t=5$ au) wave function $\psi_{\text{ref}}\left(q\right)$,
(c) initial and final $\psi_{\text{ref}}\left(p\right)$.\label{fig:morse_position_space}}
\end{figure}
\begin{figure*}
\begin{tabular}{ccc}
\includegraphics[width=0.33\textwidth]{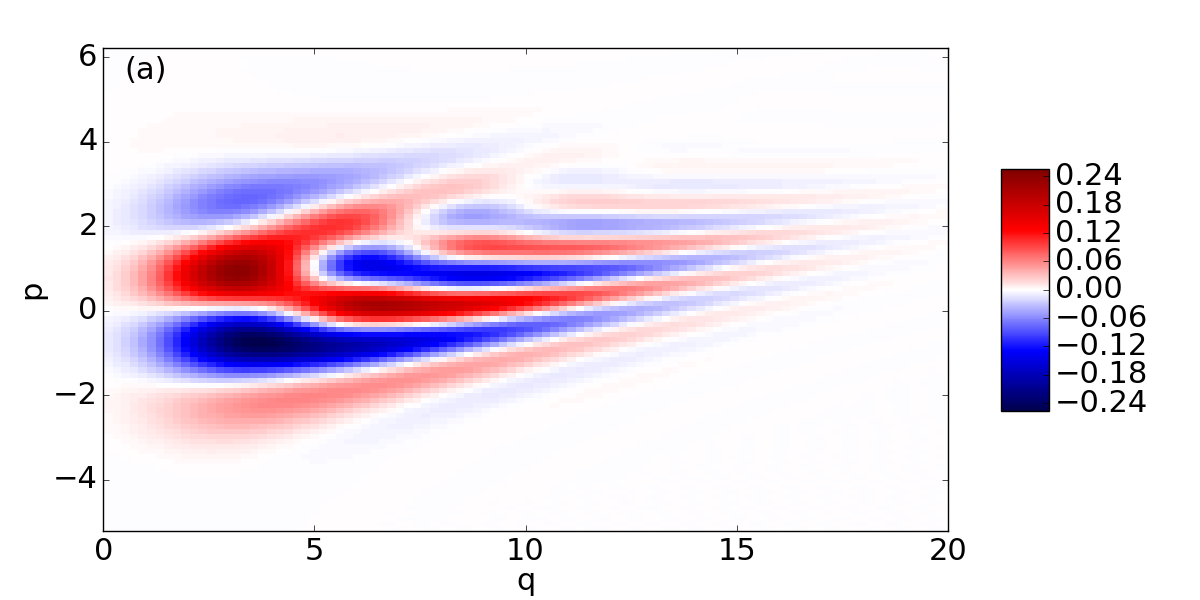} & \includegraphics[width=0.33\textwidth]{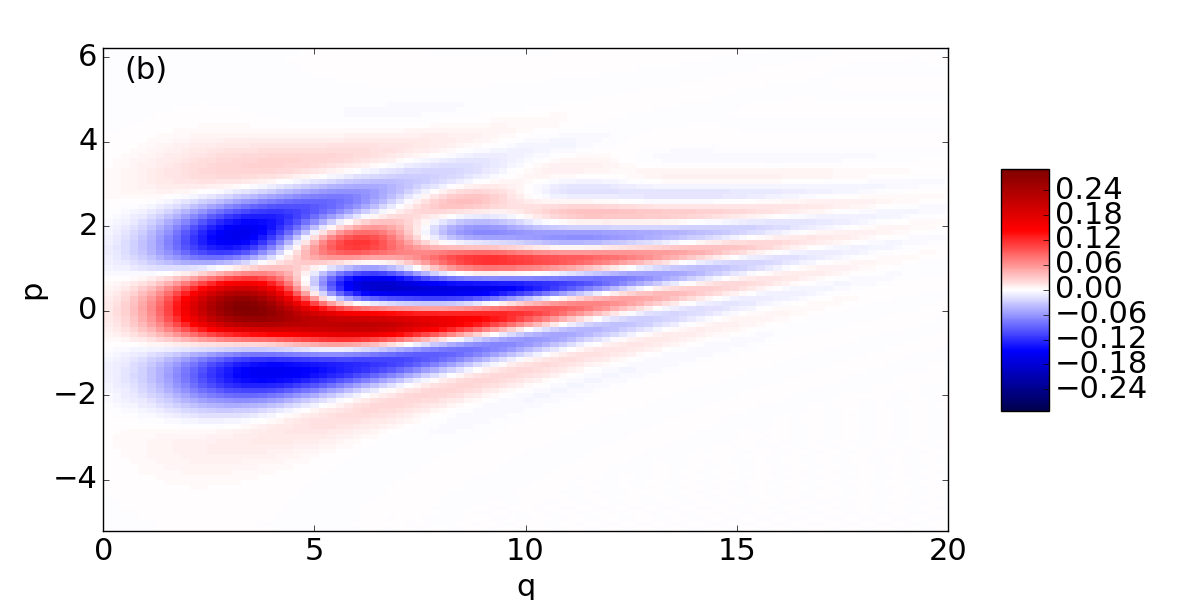} & \includegraphics[width=0.33\textwidth]{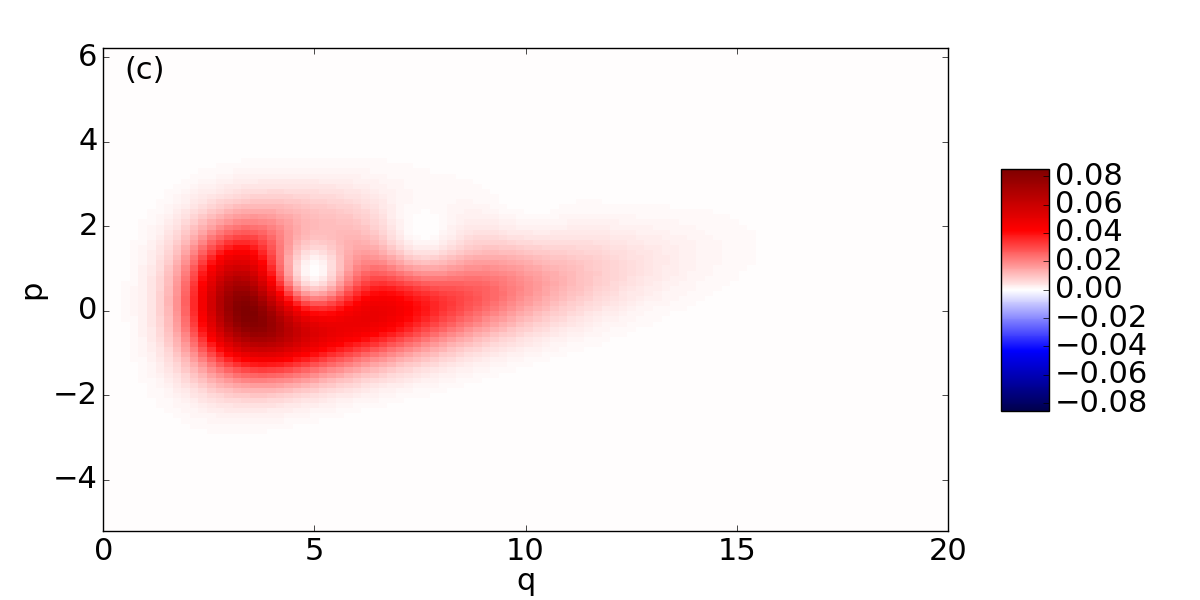}\tabularnewline
\includegraphics[width=0.33\textwidth]{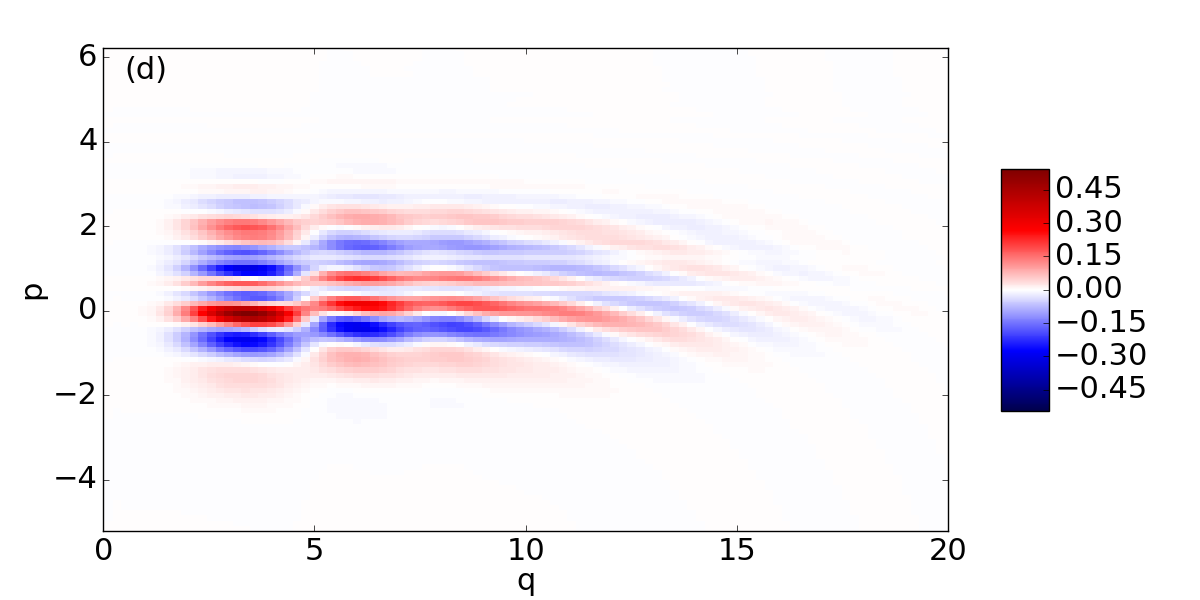} & \includegraphics[width=0.33\textwidth]{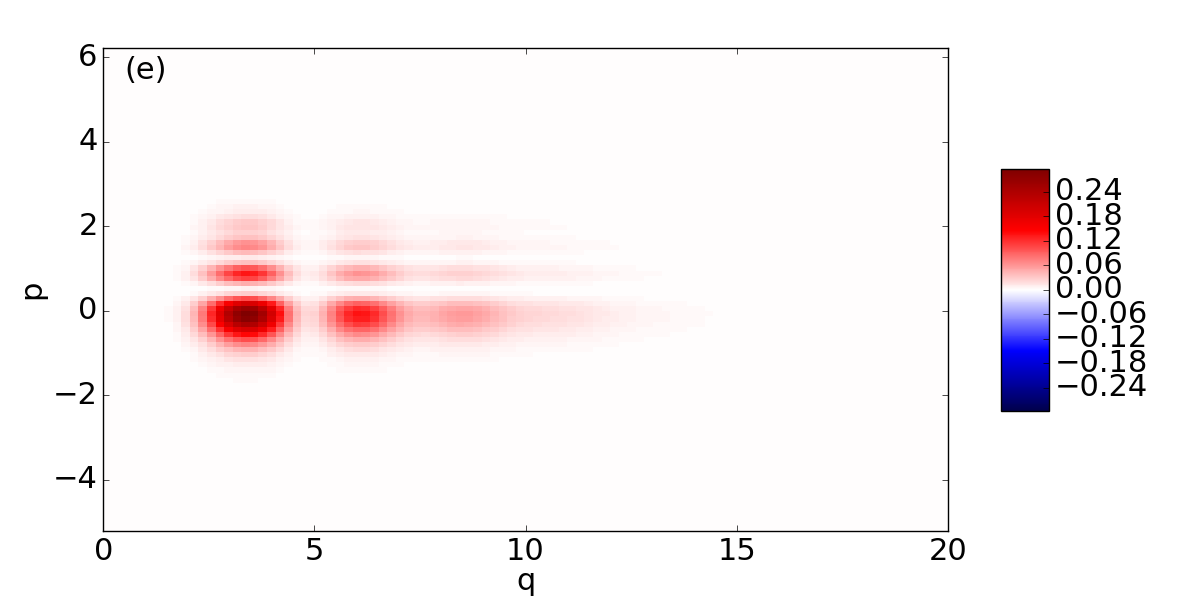} & \includegraphics[width=0.33\textwidth]{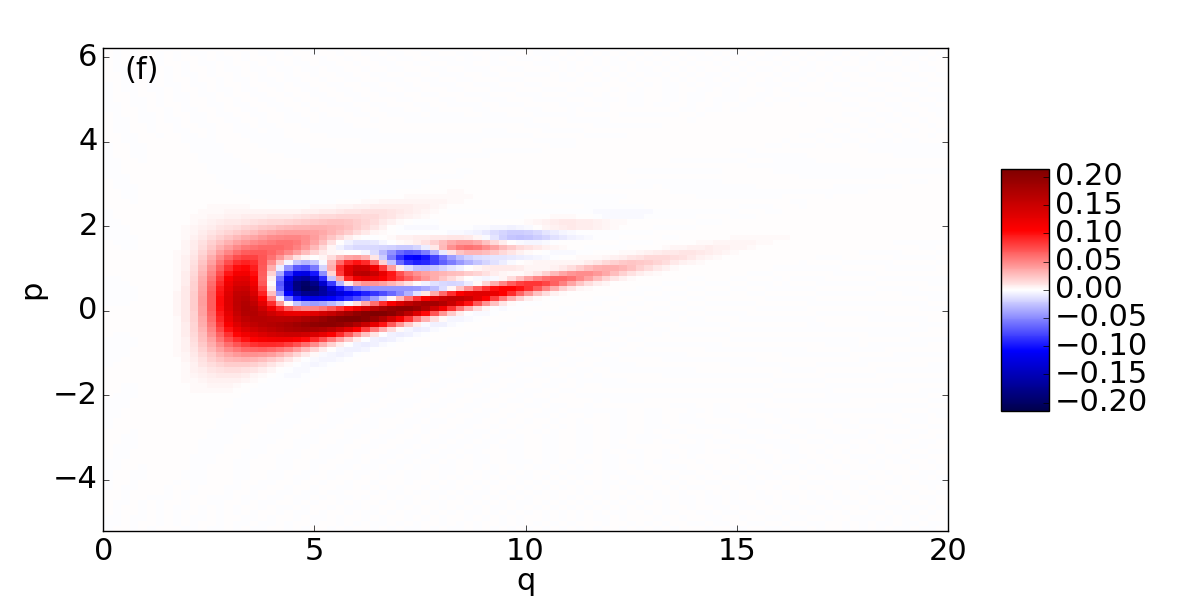}\tabularnewline
\end{tabular}

\caption{Phase-space representations of $\psi$ and $\rho$ after the propagation
for $t=5$ au. (a) $\text{Re}\left[\qp\left(q,p\right)\right]$, (b)
$\text{Im}\left[\qp\left(q,p\right)\right]$, (c) $\left|\qp\left(q,p\right)\right|^{2}$,
(d) $\text{Re}\left[P\left(q,p\right)\right]$, (e) $\left|P\left(q,p\right)\right|^{2}$,
and (f) $W\left(q,p\right)$. $\left|\qp\left(q,p\right)\right|^{2}$
clearly shows the wave packet rebounding from the repulsive part of
the Morse potential, with a small fraction finishing an oscillation,
forming the ring-like pattern in the phase-space. \label{fig:morse_phase_space}}
\end{figure*}
To numerically demonstrate properties of $\qp$ and the validity of
Eq.~(\ref{eq:prop}), the coherent state $z\left(q,-7.0,-0.5\right)$
was propagated on the Morse potential \cite{Morse1929}
\begin{equation}
V\left(q\right)=V_{0}+D\left(1-\exp\left[-A\left(q-Q\right)\right]\right)^{2}\label{eq:Morse}
\end{equation}
with parameters $V_{0}=1$, $D=0.1$, $A=0.77$, and $Q=4.0$. Figure~(\ref{fig:morse_position_space})
shows the potential and the reference wave function $\psi_{\text{ref}}$
in the position and momentum representation whereas Fig.~\ref{fig:morse_phase_space}
shows phase-space representations of the wave packet after $5$ au
of time evolution. The phase-space wave function $\qp$ was propagated
using Eq.~(\ref{eq:prop}), on a phase-space grid with $256\times256$
points in the $q$-interval $\left(-2,50\right)$ and $p$-interval
$\left(-15.4663,15.4663\right)$ using the Cash-Karp variant of the
adaptive-step Runge-Kutta algorithm \cite{Cash1990}. The reference
wave function $\psi_{\text{ref}}$ was propagated according to Eq.~(\ref{eq:tdse})
on the same $q$-grid also with the Cash-Karp algorithm. The second
derivatives of $\psi_{\text{ref}}$ with respect to positions were
computed in the momentum-space. It was verified that all wave-functions
and density matrices computed from $\qp\left(q,p,t\right)$ according
to Eqs.~(\ref{eq:z2p}), (\ref{eq:z2psi_q}), (\ref{eq:z2psi_p}),
and (\ref{eq:z2w}), i.e., $P\left(q,p,t\right)$, $\psi\left(q,t\right)$,
$\psi\left(p,t\right)$, and $W\left(q,p,t\right)$, agree with the
quantities computed from $\psi_{\text{ref}}$ within the numerical
precision of the propagation algorithm.

\bibliographystyle{apsrev4-1}
%
\onecolumngrid
\appendix

\section{transform from $\psi\left(q\right)$ to $\qp\left(q,p\right)$ }

Substituting the transformation (7) which is used to compute $\qp\left(q,p\right)$
from $\psi\left(q\right)$ into the backward transformation (5) we
get
\begin{equation}
\psi\left(q\right)=\left(2\pi\right)^{-D}\iint d^{D}p'd^{D}q'\,e^{\frac{-\left(q-q'\right)^{2}}{2}}\psi\left(q'\right)e^{i\frac{q\cdot p'}{2}}e^{-iq'\cdot p'}e^{i\frac{q\cdot p'}{2}}.\label{eq:psi_q_back_subst}
\end{equation}
Integration over $p'$ gives
\begin{equation}
\psi\left(q\right)=\int d^{D}q'\,\delta\left(q'-q\right)e^{\frac{-\left(q-q'\right)^{2}}{2}}\psi\left(q'\right),\label{eq:psi_q_back_subst_pp_int}
\end{equation}
where $\delta(q'-q)=\left(2\pi\right)^{-D}\int d^{D}p'\,e^{-i\left(q'-q\right)p'}$.
Integrating over $q'$, the validity of (7) is proven. Proofs for
transformations (4) and (6) are done in an analogous fashion.

\section{Norm of the trace of $\rhoqp\left(q,p,q',p'\right)$ }

Substituting the transform (7) into the expression for the phase-space
trace of the phase-space density matrix $\rhoqp$ (9) we obtain
\begin{eqnarray}
1 & = & \left(2\pi^{\frac{3}{2}}\right)^{-D}\iiiint d^{D}q'd^{D}p'd^{D}q''d^{D}q'''\,\nonumber \\
 &  & e^{\frac{-\left(q'-q''\right)^{2}}{2}}\psi^{*}\left(q''\right)e^{-i\frac{q'\cdot p'}{2}}e^{iq''\cdot p'}e^{\frac{-\left(q'-q'''\right)^{2}}{2}}\psi\left(q'''\right)e^{i\frac{q'\cdot p'}{2}}e^{-iq'''\cdot p'}\label{eq:norm_expanded}
\end{eqnarray}
Integrating over $p'$ gives
\begin{equation}
1=\pi^{-\frac{D}{2}}\iiint d^{D}q'd^{D}q''d^{D}q'''\,\delta\left(q'''-q''\right)e^{\frac{-\left(q'-q''\right)^{2}-\left(q'-q'''\right)^{2}}{2}}\psi^{*}\left(q''\right)\psi\left(q'''\right).\label{eq:norm_expanded_ppint}
\end{equation}
Integrating over $q'''$ leads to
\begin{equation}
1=\pi^{-\frac{D}{2}}\iint d^{D}q'd^{D}q''\,e^{-\left(q'-q''\right)^{2}}\psi^{*}\left(q''\right)\psi\left(q''\right).\label{eq:norm_expanded_qpppint}
\end{equation}
Substituting $q_{d}=q'-q''$ and $q_{a}=\frac{q'+q''}{2}$ we get
\begin{equation}
1=\iint d^{D}q_{a}\psi^{*}\left(q_{a}-\frac{q_{d}}{2}\right)\psi\left(q_{a}-\frac{q_{d}}{2}\right).\label{eq:norm_reduced2q}
\end{equation}
Since $\psi$ is normalized to unity, Eq.~(9) is proven.

\section{Phase-space identity transform}

Substituting Eq.~(7) into the phase-space identity transform (10)
we obtain
\begin{equation}
\qp\left(q,p\right)=\left(4\pi\right)^{-D}\left(2\pi^{\frac{3}{2}}\right)^{-\frac{D}{2}}\iiint d^{D}q'd^{D}p'd^{D}q''\,e^{\frac{-\left(q'-q''\right)^{2}}{2}}\psi\left(q''\right)e^{i\frac{q'\cdot p'}{2}}e^{-iq''\cdot p'}e^{i\frac{q\cdot p'-q'\cdot p}{2}}.\label{eq:identity2q}
\end{equation}
Integrating over $p'$ gives
\begin{equation}
\qp\left(q,p\right)=2^{-D}\left(2\pi^{\frac{3}{2}}\right)^{-\frac{D}{2}}\iint d^{D}q'd^{D}q''\,\delta\left(\frac{q'+q-2q''}{2}\right)e^{\frac{-\left(q'-q''\right)^{2}}{2}}\psi\left(q''\right)e^{-i\frac{q'\cdot p}{2}}.\label{eq:identity2q_pp_int}
\end{equation}
Integrating over $q''$ leads to
\begin{equation}
\qp\left(q,p\right)=2^{-D}\left(2\pi^{\frac{3}{2}}\right)^{-\frac{D}{2}}\int d^{D}q'\,e^{\frac{-\left(\frac{q'-q}{2}\right)^{2}}{2}}\psi\left(\frac{q'+q}{2}\right)e^{-i\frac{q'\cdot p}{2}}.\label{eq:identity2q_qpp_int}
\end{equation}
Finally, substituting $q_{a}=\frac{q'+q}{2}$ one recovers Eq.~(7)

\begin{equation}
\qp\left(q,p\right)=\left(2\pi^{\frac{3}{2}}\right)^{-\frac{D}{2}}\int d^{D}q_{a}\,e^{\frac{-\left(q_{a}-q\right)^{2}}{2}}\psi\left(q_{a}\right)e^{-i\frac{\left(2q_{a}-q\right)\cdot p}{2}}.\label{eq:identity2q_eq}
\end{equation}

\section{Equation of motion}

In order to prove that Eq.~(13) reduces to Schr\"{o}dinger equation,
the Hamiltonian $\hat{\mathbf{H}}=\hat{\mathbf{T}}+\hat{\mathbf{V}}$
is first split into the kinetic and potential energy part. Writing
Eq.~(13) for the potential energy $\hat{\mathbf{V}}$ and using Eq.~(7)
to express $\qp$ we obtain
\begin{equation}
\hat{\mathbf{V}}\qp=\left(4\pi\right)^{-D}\left(2\pi^{\frac{3}{2}}\right)^{-\frac{D}{2}}\iiint d^{D}q'd^{D}p'd^{D}q''\,V\left(\frac{q'+q}{2}\right)e^{\frac{-\left(q'-q''\right)^{2}}{2}}\psi\left(q''\right)e^{i\frac{q'\cdot p'}{2}}e^{-iq''\cdot p'}e^{i\frac{q\cdot p'-q'\cdot p}{2}}.\label{eq:Vqp_full}
\end{equation}
Integrating over $p'$ gives
\begin{equation}
\hat{\mathbf{V}}\qp=2^{-D}\left(2\pi^{\frac{3}{2}}\right)^{-\frac{D}{2}}\iint d^{D}q'd^{D}q''\,\delta\left(\frac{q'+q-2q''}{2}\right)V\left(\frac{q'+q}{2}\right)e^{\frac{-\left(q'-q''\right)^{2}}{2}}\psi\left(q''\right)e^{-i\frac{q'\cdot p}{2}}.\label{eq:Vqp_int_p'}
\end{equation}
Integrating over $q''$ then leads to
\begin{equation}
\hat{\mathbf{V}}\qp=2^{-D}\left(2\pi^{\frac{3}{2}}\right)^{-\frac{D}{2}}\int d^{D}q'\,e^{\frac{-\left(\frac{q'-q}{2}\right)^{2}}{2}}V\left(\frac{q'+q}{2}\right)\psi\left(\frac{q'+q}{2}\right)e^{-i\frac{q'\cdot p}{2}}.\label{eq:Vqp_int_q'}
\end{equation}
Substituting $q_{a}=\frac{q'+q}{2}$,

\begin{equation}
\hat{\mathbf{V}}\qp=\left(2\pi^{\frac{3}{2}}\right)^{-\frac{D}{2}}\int d^{D}q_{a}\,e^{\frac{-\left(q_{a}-q\right)^{2}}{2}}V\left(q_{a}\right)\psi\left(q_{a}\right)e^{-i\frac{\left(2q_{a}-q\right)\cdot p}{2}},\label{eq:Vqp_subst}
\end{equation}
which is nothing else than transform (7) of the position-space product
of $V\left(q_{a}\right)$ with the wave function $\psi\left(q_{a}\right)$.

Analogously, for the kinetic energy part $\hat{\mathbf{T}}$ we obtain
\begin{equation}
\hat{\mathbf{T}}\qp=\left(4\pi\right)^{-D}\left(2\pi^{\frac{3}{2}}\right)^{-\frac{D}{2}}\iiint d^{D}q'd^{D}p'd^{D}p''\,T\left(\frac{p'+p}{2}\right)e^{\frac{-\left(p'-p''\right)^{2}}{2}}\psi\left(p''\right)e^{-i\frac{q'\cdot p'}{2}}e^{iq'\cdot p''}e^{i\frac{q\cdot p'-q'\cdot p}{2}},\label{eq:Tqp_full}
\end{equation}
where the momentum-space version of Eq.~(7) is used

\begin{equation}
\qp\left(q,p\right)=\left(2\pi^{\frac{3}{2}}\right)^{-\frac{D}{2}}e^{-i\frac{q\cdot p}{2}}\int d^{D}p'\,e^{\frac{-\left(p-p'\right)^{2}}{2}}\psi\left(p'\right)e^{iq\cdot p'}.\label{eq:psi_p2z}
\end{equation}
Integrating over $q'$ gives
\begin{equation}
\hat{\mathbf{T}}\qp=2^{-D}\left(2\pi^{\frac{3}{2}}\right)^{-\frac{D}{2}}\iint d^{D}p'd^{D}p''\,\delta\left(\frac{p'+p-2p''}{2}\right)T\left(\frac{p'+p}{2}\right)e^{\frac{-\left(p'-p''\right)^{2}}{2}}\psi\left(p''\right)e^{i\frac{q\cdot p'}{2}}.\label{eq:Tqp_int_q'}
\end{equation}
Integrating over $p''$
\begin{equation}
\hat{\mathbf{T}}\qp=2^{-D}\left(2\pi^{\frac{3}{2}}\right)^{-\frac{D}{2}}\int d^{D}p'\,e^{\frac{-\left(\frac{p'-p}{2}\right)^{2}}{2}}T\left(\frac{q'+q}{2}\right)\psi\left(\frac{p'+p}{2}\right)e^{i\frac{q\cdot p'}{2}}.\label{eq:Tqp_int_p''}
\end{equation}
Substituting $p_{a}=\frac{p'+p}{2}$ one recovers the momentum-space
version of Eq.~(7)

\begin{equation}
\hat{\mathbf{T}}\qp=\left(2\pi^{\frac{3}{2}}\right)^{-\frac{D}{2}}\int d^{D}p_{a}\,e^{\frac{-\left(p_{a}-p\right)^{2}}{2}}T\left(p_{a}\right)\psi\left(p_{a}\right)e^{i\frac{\left(2p_{a}-p\right)\cdot q}{2}}\label{eq:Tqp_subst}
\end{equation}
for the product of $T\left(p_{a}\right)$ with $\psi\left(p_{a}\right)$. 

\section{Mean values}

Substituting into Eq.~(16), the mean value of the operator which
is a function of the position only can be written as 
\begin{eqnarray}
\left\langle \hat{\mathbf{O}}_{q}\right\rangle  & = & \left(4\pi\right)^{-D}\left(2\pi^{\frac{3}{2}}\right)^{-D}\dotsint d^{D}qd^{D}pd^{D}q'''d^{D}q'd^{D}p'd^{D}q''\,\nonumber \\
 &  & e^{\frac{-\left(q-q'''\right)^{2}}{2}}\psi^{*}\left(q'''\right)e^{-i\frac{q\cdot p}{2}}e^{iq'''\cdot p}O\left(\frac{q'+q}{2}\right)e^{\frac{-\left(q'-q''\right)^{2}}{2}}\psi\left(q''\right)e^{i\frac{q'\cdot p'}{2}}e^{-iq''\cdot p'}e^{i\frac{q\cdot p'-q'\cdot p}{2}},\label{eq:o_q_full}
\end{eqnarray}
where Eq.~(7) was used to express both $\qp^{*}\left(q,p\right)$
and $\qp\left(q',p'\right)$ in terms of position-space wave functions
$\psi\left(q\right)$. Integrating over $p$ and $p'$ gives

\begin{eqnarray}
\left\langle \hat{\mathbf{O}}_{q}\right\rangle  & = & \left(2\sqrt{\pi}\right)^{-D}\iiiint d^{D}qd^{D}q'''d^{D}q'd^{D}q''\,\nonumber \\
 &  & e^{\frac{-\left(q-q'''\right)^{2}}{2}}\psi^{*}\left(q'''\right)\delta\left(\frac{2q'''-q'-q}{2}\right)\delta\left(\frac{q'+q-2q''}{2}\right)O\left(\frac{q'+q}{2}\right)e^{\frac{-\left(q'-q''\right)^{2}}{2}}\psi\left(q''\right).\label{eq:o_q_int_p_p'}
\end{eqnarray}
Integrating over $q''$ and $q'''$ leads to
\begin{equation}
\left\langle \hat{\mathbf{O}}_{q}\right\rangle =\left(2\sqrt{\pi}\right)^{-D}\iint d^{D}qd^{D}q'\,e^{\frac{-\left(\frac{q-q'}{2}\right)^{2}}{2}}e^{\frac{-\left(\frac{q'-q}{2}\right)^{2}}{2}}\psi^{*}\left(\frac{q'+q}{2}\right)V\left(\frac{q'+q}{2}\right)\psi\left(\frac{q'+q}{2}\right).\label{eq:o_q_int_q''_q'''}
\end{equation}
Substituting $q_{a}=\frac{q'+q}{2}$ and $q_{b}=\frac{q'-q}{2}$ simplifies
the formula to

\begin{equation}
\left\langle \hat{\mathbf{O}}_{q}\right\rangle =\pi^{-\frac{D}{2}}\iint d^{D}q_{a}d^{D}q_{b}\,e^{-q_{b}^{2}}\psi^{*}\left(q_{a}\right)V\left(q_{a}\right)\psi\left(q_{a}\right).\label{eq:o_q_subst}
\end{equation}
Finally, integrating over $q_{b}$, the usual position-space expression
for the mean value is recovered

\begin{equation}
\left\langle \hat{\mathbf{O}}_{q}\right\rangle =\int d^{D}q_{a}\,\psi^{*}\left(q_{a}\right)V\left(q_{a}\right)\psi\left(q_{a}\right).\label{eq:mean_position_space}
\end{equation}
The expression for the mean value of the momentum only operator $\left\langle \hat{\mathbf{O}}_{p}\right\rangle $
is proven analogously in the momentum-space using the momentum-space
transforms (\ref{eq:psi_p2z}).

\end{document}